\documentclass[preprint,12pt]{elsarticle}
\usepackage{amssymb}
\usepackage{multirow}
\usepackage{graphicx}
\usepackage{subfigure}
\usepackage{epsfig}
\usepackage[figuresright]{rotating}
\usepackage[dvips]{color}
\journal{Chemical Physics Letters}

\begin{document}

\begin{frontmatter}
\title{Structural and vibrational properties of nitrogen-rich energetic material guanidinium 2-methyl-5-nitraminotetrazolate}
\author{K. Ramesh Babu and G. Vaitheeswaran}
\address{Advanced Centre of Research in High Energy Materials (ACRHEM),
University of Hyderabad, Prof. C. R. Rao Road, Gachibowli, Andhra Pradesh, Hyderabad- 500 046, India\\
\vspace{0.6in}
*Corresponding author E-mail address: gvsp@uohyd.ernet.in\\
\hspace {1.2in} Tel No.: +91-40-23138709\\
\hspace {1.2in} Fax No.: +91-40-23010227}
\end{frontmatter}
\newpage
\section*{Abstract} We present density functional theory calculations on the crystal structure, equation of state, vibrational properties and electronic structure of nitrogen-rich solid energetic material guanidinium 2-methyl-5-nitraminotetrazolate (G-MNAT). The ground state structural properties calculated with dispersion corrected density functionals are in good agreement with experiment. The computed equilibrium crystal structure is further used to calculate the equation of state and zone-center vibrational frequencies of the material. The electronic band structure is calculated and found that the material is an indirect band gap semiconductor with a value of 3.04 eV.\\
Keywords:
green energetic materials; van der Waals interactions; equation of state; vibrational frequencies; electronic structure
\newpage
\section{Introduction}
Energetic materials have wide range of applications in both civilian and military sectors. For example, energetic materials are used in agriculture, mining, building demolition, pyrotechnics, coal blasting, tunneling, welding etc. Most importantly they are useful in fire fighting in war heads and as rocket boosters in aero space applications \cite{Agrawal, Klapotke, Akhavan, Fair}. In general, energetic materials contain both fuel and oxidizer and reacts readily with the release of huge chemical energy stored within their molecular structures upon external stimulus such as heat, impact, shock, electric spark etc. The amount of energy released from an energetic material within a short time is considerably large when compared to the normal materials \cite{Talawar}.
\paragraph*{} The detonation products of most of the energetic materials are water vapour, carbon monoxide CO, and carbon-di-oxide CO$_2$ \cite{Talawar, Sikder}. These are well known green house gases that greatly affect the temperature of the earth. The best remedy for this problem is to have energetic materials that only give environmentally clean and eco-friendly detonation products. Nitrogen-rich compounds meet these demands quite well as they tend to show a high energy content and most importantly, their detonation products are pure nitrogen gas which is environment friendly \cite{Sikder}. Eremets et al., have synthesized polymeric nitrogen which is considered to be a green high energy density material \cite{Eremets}. Very recently, Fendt et al., synthesized tetrazole based nitrogen-rich energetic materials whose performance characteristics are found to be in good accord with well known high explosives \cite{Fendt}. In this family of compounds, guanidinium 2-methyl-5-nitraminotetrazolate (C$_3$H$_9$N$_9$O$_2$, G-MNAT) receives particular interest because of its similar energetic characteristics with that of 1,3,5-Trinitroperhydro-1,3,5-triazine (C$_3$H$_6$N$_6$O$_6$, RDX).
\paragraph*{}
For any energetic material, the physical and chemical properties such as electronic band structure, bonding and vibrational properties are very important in order to understand the stability and thereby the sensitivity of the materials. These properties are directly related to the molecular packing, symmetry of the crystal structure and most importantly to the crystal density. Therefore it is quite essential to know about the crystal structure of the energetic materials and the structural modifications that occur upon the application of high pressures. Density Functional Theory (DFT) is a successful tool in simulating and predicting the physical and chemical properties of a wide spectrum of energetic materials \cite{Oleyn, YMGupta, Perger, Byrd, Chaba}. However, most of the energetic materials have complicated crystal structures with weak inter molecular interactions and hence the investigation of different physical and chemical properties of energetic materials through DFT is really a challenging task \cite{Rice}. In recent years, the advances in DFT methods have enabled us to describe effectively the materials with weak dispersive interactions \cite{Grimme, TS}. There are few theoretical studies available based on dispersion corrected density functional methods to study the structural properties of various energetic materials \cite{Rice}. To the best of our knowledge there are no theoretical reports available on solid G-MNAT. In this present work, we aim to study the crystal structure, equation of state and vibrational properties of solid G-MNAT. It is a well known fact that the electronic band gap plays a major role to understand the sensitivity of energetic materials \cite{Xiao}. Hence, we also calculate the energy band structure and the variation of band gap with pressure. The remainder of the paper is organized as follows: A brief description of our computational details is presented in section 2. The results and discussion are presented in section 3 followed by summary of our conclusions in section 4.
\section{Computational details}
The calculations are performed using plane wave pseudopotential method based on density functional theory \cite{Segall}. The interactions between the ions and electrons are described by using Vanderbilt ultrasoft pseudopotentials \cite{Vanderbilt}. For all the calculations we have included the 1s$^1$ electrons for hydrogen, 2s$^2$, 2p$^2$ electrons for carbon, 2s$^2$, 2p$^3$ electrons of nitrogen and the 2s$^2$, 2p$^4$ states of oxygen. Both local density approximation (LDA) of Ceperley and Alder \cite{Ceperley} parameterized by Perdew and Zunger (CA-PZ) \cite{PPerdew} and also the generalized gradient approximation (GGA) with the Perdew-Burke-Ernzerhof (PBE) \cite{Perdew} parameterization are used for the exchange-correlation potentials. The calculations are performed using an energy cut-off of 640 eV for the plane wave basis set. Integrations in the Brillouin zone are performed according with a 5x3x2 Monkhorst-Pack grid scheme \cite{Monkhorst} k-point mesh. The changes in the total energies with the number of k-points and the cut-off energy are tested to ensure the convergence within one meV per atom.
\paragraph*{} To treat van der Waals (vdW) interactions efficiently, we have used the vdW correction to the exchange - correlation functional of standard density functional theory at semi-empirical level. According to semi-empirical dispersion correction approach, the total energy of the system can be expressed as
\begin{equation}
E_{total} = E_{DFT} + E_{Disp}
\end{equation}
where
\begin{equation}
E_{Disp} = s_i\Sigma_{i=1}^N\Sigma_{j>i}^Nf(S_RR^{0}_{ij}, R_{ij})C_{6, ij}R_{ij}^{-6}
\end{equation}
here C$_{6, ij}$ is called dispersion coefficient between any atom pair $i$ and $j$ which solely depends upon the material and R$_{ij}$ is the distance between the atoms $i$ and $j$. In the present study we have used the recently developed dispersion schemes by Grimme \cite{Grimme}, Tkatchenko - Scheffler \cite{TS} within GGA. These semi-empirical approaches provide the best compromise between the cost of first principles evaluation of the dispersion terms and the need to improve non-bonding interactions in the standard DFT description.
\section{Results and discussion}
\subsection{Structural properties}
At ambient pressure, solid G-MNAT exists in monoclinic structure with space group P2$_1$ and contain two molecules per unit cell (z=2) \cite{Fendt}. The experimental crystal structure is taken as starting input for the calculations and then we apply standard density functionals LDA (CA-PZ), GGA (PBE) and also the dispersion corrected density functionals PBE+TS and PBE+G06 to get the theoretical equilibrium structure. The calculated lattice parameters are presented in Table 1 together with the experimental values \cite{Fendt}. The computed crystal volume is underestimated by -8.3$\%$ using CA-PZ functional and overestimated by 7.1$\%$ with PBE computation. This large discrepancy is due to the fact that the present studied compound is a molecular solid with weak dispersion forces for which the usual LDA and GGA functionals are inadequate to treat these forces. On the other hand, the computation carried out by dispersion corrected density functionals describe well the crystal structure and the computed volume is in good agreement with experiment. The equilibrium crystal volume is overestimated by 1.7$\%$ using PBE+TS functional and underestimated by -0.6$\%$ with PBE+G06 functional. Clearly, the dispersion corrected density functionals are efficient to describe the crystal structure of G-MNAT. This is supported by the earlier theoretical studies on energetic solids where the authors found that the intermolecular interactions were well described by the  dispersion corrected density functionals \cite{Rice}. In particular, the PBE+G06 functional describes the solid G-MNAT system with less error compared to PBE+TS functional. Thus, for all rest of the calculations we use PBE+G06 functional.
\paragraph*{}The optimized crystal structure of G-MNAT and its molecular structure using PBE+G06 functional are shown in Figure 1(a) and 1(b) respectively. The calculated bond lengths between N1-N2, N2-N3, and N3-N4 are found to be 1.35 {\AA}, 1.34 {\AA} and 1.33 {\AA} respectively. These are in good agreement with experimental bond length values that varies between 1.30-1.33 {\AA} \cite{Fendt}. The torsion angle C1-N2-N3-N4 is calculated to be -0.246$^0$ which is in good comparison with experimental value of -0.3$^0$ \cite{Fendt}. The bond lengths between the C1-N1, C1-N4 and C2-N2 are calculated to be 1.36 {\AA}, 1.37 {\AA}, 1.45 {\AA}. It should be note that the calculated values of bond lengths between N atoms and C, N atoms are comparable to that of the bond lengths of N-N (1.45 {\AA}), N=N(1.25 {\AA}), C-N (1.47 {\AA}) and C=N (1.22 {\AA}) respectively.
\paragraph*{}
The equation of state for G-MNAT crystal is obtained by performing the hydrostatic compression simulation and shown in Figure 2(a). In practice, the experiments on energetic solids are limited to low pressures (less than 10 GPa) hence we confined the pressure limit to 10 GPa. From the figure 2(a), it can be seen that the volume decreases monotonically with V/V$_0$ = 77$\%$ at 10 GPa. The lattice constants as a function of pressure are shown in Figure 2(b). Among the three lattice parameters, the reduction in lattice parameter `a' with pressure is found to be more when compared to other with a/a$_0$ = 86$\%$ while b/b$_0$ = 98$\%$ and c/c$_0$ = 89$\%$. This implies that the G-MNAT lattice is much stiffer in b-direction than a- and c-directions. This can be interpreted from the perspective view of crystalline G-MNAT along three directions as shown in Figure 1(a). In G-MNAT crystal, the intermolecular distance along the a-axis is largest and therefore the interactions between the molecules are relatively weak which results in high compressibility along a-direction compared to other. From this we conclude that the compressibility of G-MNAT lattice is anisotropic. In Figure 2(c), the variation of the crystal density, $\rho$ (gm/cc) with pressure is shown. Clearly, $\rho$ increases with pressure (as volume decreases) which indicates that the intermolecular interactions enhance under pressure and thereby the lattice become more stiffer at high pressure. The monoclinic angle decreases with pressure to a value of 84.02$^0$ at 10 GPa as shown in Figure 2(d).
\paragraph*{}The calculated equation of state of solid G-MNAT can be used to obtain the bulk modulus B$_0$ and its pressure derivative B$_0$$^\prime$ by fitting the pressure-volume data to the Murnaghan's equation of state. The calculated values of B$_0$ and B$_0$$^\prime$ are found to be 15.3 GPa and 4.66 respectively and are presented Table 1. It can be noticed that when compared to the bulk modulus value of other high energetic solids, for example, PETN (B$_0$ = 9.1 GPa) \cite{Conroy}, RDX (B$_0$ = 11.9 GPa) \cite{Stevens}, HMX (B$_0$ = 12.5 GPa) \cite{Chen} and TAG-MNT (B$_0$ = 14.6 GPa) \cite{McWilliams}, G-MNAT is a stiffer material.
\subsection{Electronic band structure}
The electronic band structure calculated for the ground state equilibrium structure of solid G-MNAT with PBE+G06 functional is displayed in Figure 3(a). The band structure is plotted along the high symmetry points of the Brillouin zone of monoclinic lattice. From the band structure we conclude that solid G-MNAT is an indirect band gap semiconductor with a gap of 3.04 eV occur between the Z-point of the valence band and A-point of conduction band. The variation of band gap as a function of pressure is shown in Figure 3(b). Clearly, the magnitude of the band gap decreases with pressure and reaches to 2.55 eV at 10 GPa. In the case of energetic materials, it was shown that there is a correlation between the band gap and sensitivity of the material through `principle of easiest transition' (PET) \cite{Xiao}. According to PET, for energetic crystals with smaller band gap value it is easier the electron transfer from valence band to the conduction band and hence become more decomposed and exploded. By using this criterion, the experimental sensitivity order of metal azides, high energetic materials such as polymorphs of HMX, CL-20 is successfully explained \cite{Xiao}. Therefore, in this present study we try to correlate the calculated band gap to the sensitivity of G-MNAT crystal. As pressure increases the band gap value decreases indicating that the sensitivity of G-MNAT crystal increases with pressure. This is supported by the fact that an applied pressure increases the sensitivity of an energetic material \cite{Xiao}.
\subsection{Vibrational properties}
The vibrational properties are calculated within the framework of density functional perturbation theory (DFPT) \cite{Gonze, Refson, Bar}. We use norm-conserving pseudopotentials for the calculation of zone-center phonon frequencies with an energy cut-off of 800 eV. The Brillouin zone integration is performed for 5x3x2 grid of k-points. The unit cell of the solid G-MNAT contains 46 atoms giving rise to a total of 138 vibrational modes. The \textit{P2$_1$} space group, which describes the monoclinic symmetry, has two irreducible representations namely, A and B. A group-theoretical analysis gives the following decomposition of vibrational representation into its irreducible components at the $\Gamma$-point: $\Gamma$-tot = 69A + 69B. Out of these modes, three (A + 2B) are acoustic modes and the remaining 135 modes are optical modes. All these modes are found to be both infrared and Raman active.
\paragraph{}
The calculated vibrational frequencies along with irreducible representation and mode assignment are presented in Table 2, Table 3 and Table 4. The vibrational modes that are situated between the frequency range from 74.2 cm$^{-1}$ to 188.2 cm$^{-1}$ are due to the collective vibrations from the cation and anion, thus these modes are designated as lattice modes and are presented in Table 2. The modes that involve vibrations of the internal molecular geometry of cation and anion are in between the frequency range from 192.8 cm$^{-1}$ to 1687.9 cm$^{-1}$ and are given in Table 2 and Table 3. The wagging vibrations of NH$_2$ cation are lying at 192.8 cm$^{-1}$, 223.5 cm$^{-1}$, 523.8 cm$^{-1}$ to 633.8 cm$^{-1}$ and at 774.2 cm$^{-1}$. The modes situated at 715.2 cm$^{-1}$ and 715.8 cm$^{-1}$ are due to the wagging of the N=C of anion. The stretching vibrations of tetrazole ring are found to present between 1011.3 cm$^{-1}$ to 1027.8 cm$^{-1}$ which are in good agreement with the experimental observed range of 1011 cm$^{-1}$ to 1026 cm$^{-1}$. The -C=N stretching vibration of the tetrazole ring is calculated to be 1687.9 cm$^{-1}$, which is in good agreement with experimental frequency of 1698 cm$^{-1}$. The other important vibrational modes are due to the C-H stretching vibrations of the methyl group attached to the tetrazole ring and N-H stretching vibrations of the amino groups of the cation.
\paragraph*{}According to experiment, the C-H stretching modes are situated at 2956 cm$^{-1}$ to 2970 cm$^{-1}$ while the symmetric and asymmetric stretching of N-H modes are observed in the range from 3238 cm$^{-1}$ to 3484 cm$^{-1}$. From the present PBE+G06 calculations, we find the frequency of stretching of C-H modes are situated between 2963.5 cm$^{-1}$ to 3196.7 cm$^{-1}$. On the other hand, symmetric stretching frequencies of N-H mode of cation are found to present between 3242.8 cm$^{-1}$ to 3279.9 cm$^{-1}$ and the asymmetric stretching N-H modes are situated at 3316.7 cm$^{-1}$ to 3483.3 cm$^{-1}$. Overall, the present study of vibrational properties based on dispersion corrected density functional PBE+G06 could reproduce well the observed vibrational frequencies of G-MNAT crystal. This confirms the fact that vdW corrected density functionals are essential to compute the structural and vibrational properties of solid energetic systems.
\paragraph*{}
\section{Conclusion}
In conclusion, we have studied crystal structure, equation of state, electronic structure and vibrational properties of solid G-MNAT. We found that PBE+G06 functional gives accurate structural results when compared to CA-PZ, PBE and PBE+TS functionals. The optimized crystal structure with the PBE+G06 functional was then used for the calculation of electronic band structure and vibrational properties. We found that the compressibility of solid G-MNAT is anisotropic and the lattice is more compressible along a-axis and least compressible along b-axis. The calculated electronic band structure show that G-MNAT is an indirect band gap semiconductor with a gap of 3.04 eV. The band gap decreases with pressure implies that the material is more sensitive under pressure. The complete zone-center frequencies are calculated and each mode was assigned according to their molecular vibration. The calculated vibrational frequencies are in good agreement with experimental data.
\section{ACKNOWLEDGMENTS}
K R B would like to acknowledges DRDO for funding through ACRHEM and also CMSD, University of Hyderabad for providing computational facility.

\newpage
\textbf{Figure Legends:} \vspace{1cm}  \\
Figure 1: (Colour Online) Crystal structure of G-MNAT (a). In figure grey ball, white ball, blue ball and red ball indicates, carbon, hydrogen, nitrogen and oxygen atoms respectively. (b) Molecular structure of G-MNAT \vspace{1cm} \\

Figure 2: (Colour online) Volume variation of solid G-MNAT with pressure (a) variation of lattice parameters with pressure (b) density as function of pressure (c) and pressure dependance of monoclinic angle (d). \vspace{1cm}\\


Figure 3: (Colour online) Electronic band structure of G-MNAT calculated within PBE+G06 functional (a). Variation of electronic band gap with pressure (b). \vspace{1cm}\\

\clearpage
%
%
\begin{figure}
\begin{center}
\subfigure[]{\includegraphics[width=80mm,height=60mm]{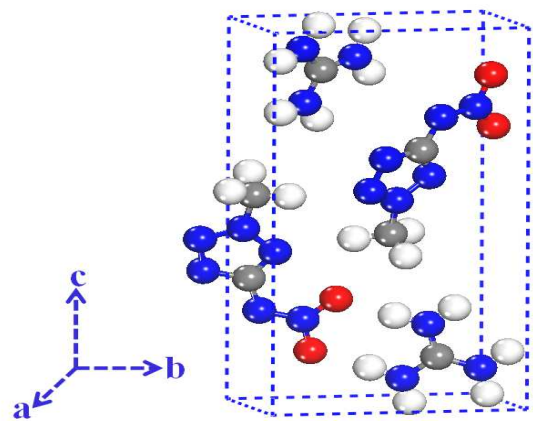}}
\subfigure[]{\includegraphics[width=65mm,height=40mm]{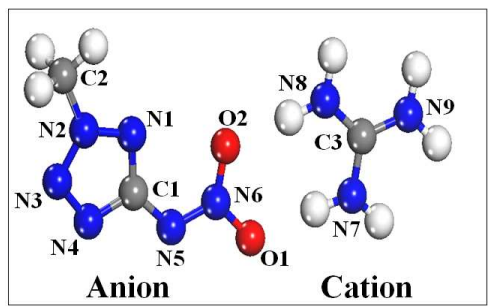}}\\
\caption{(Colour Online) Crystal structure of G-MNAT (a). In figure grey ball, white ball, blue ball and red ball indicates, carbon, hydrogen, nitrogen and oxygen atoms respectively. (b) Molecular structure of G-MNAT.}
\end{center}
\end{figure}

%

\clearpage
\newpage
\begin{figure}
\begin{center}
\subfigure[]{\includegraphics[width=60mm,height=60mm]{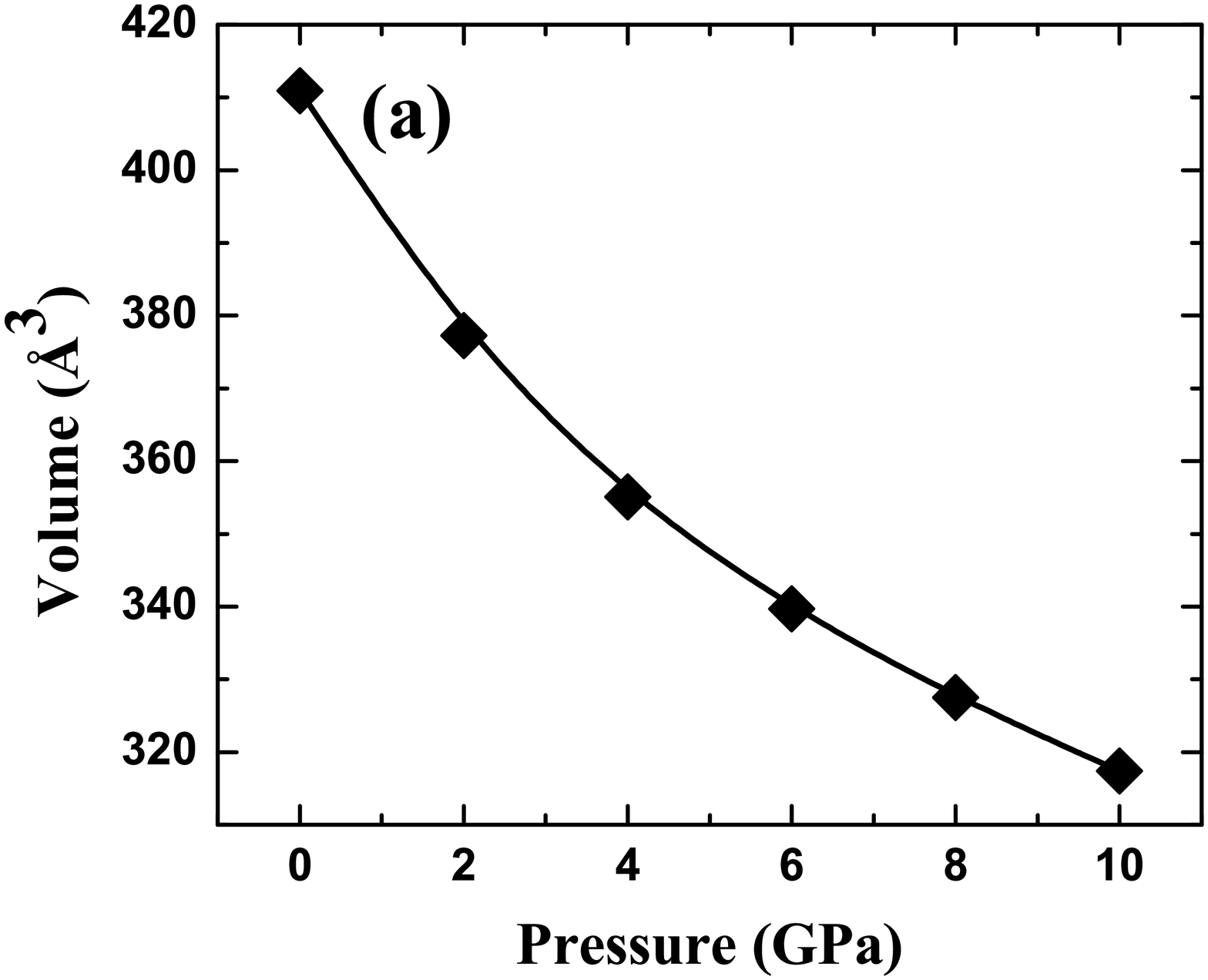}}
\subfigure[]{\includegraphics[width=60mm,height=60mm]{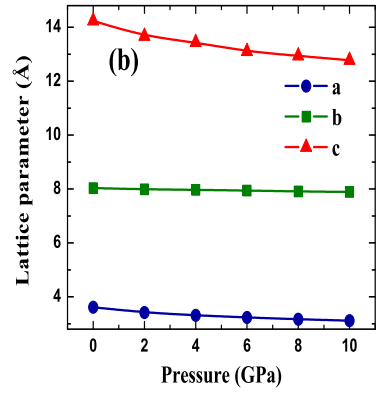}}\
\subfigure[]{\includegraphics[width=60mm,height=60mm]{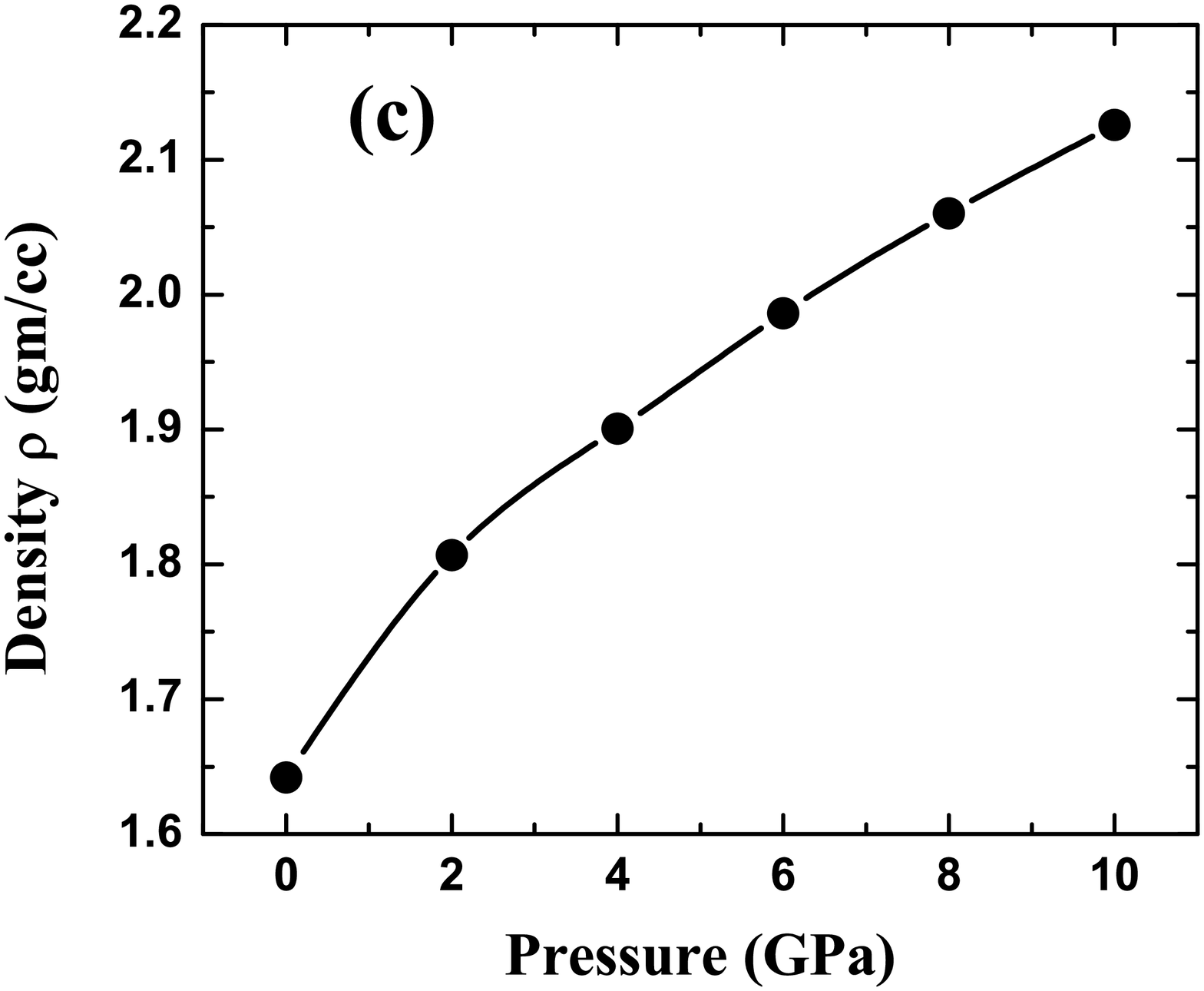}}
\subfigure[]{\includegraphics[width=60mm,height=60mm]{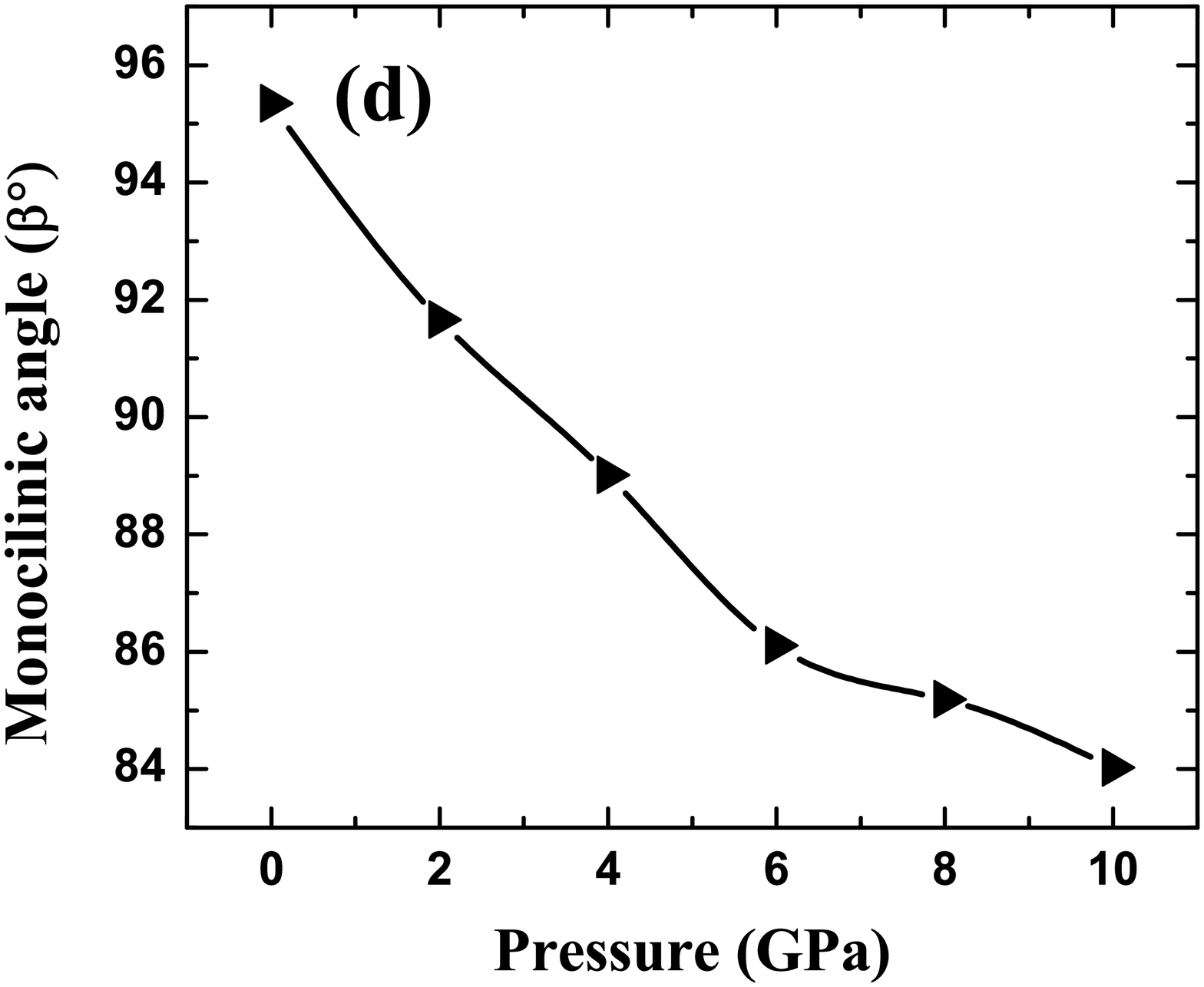}}\\
\caption{(Colour online) Volume variation of solid G-MNAT with pressure (a) variation of lattice parameters with pressure (b) density as function of pressure (c) and pressure dependance of monoclinic angle (d).}
\end{center}
\end{figure}
\clearpage
\newpage

\begin{figure}
\begin{center}
\subfigure[]{\includegraphics[width=100mm,height=100mm]{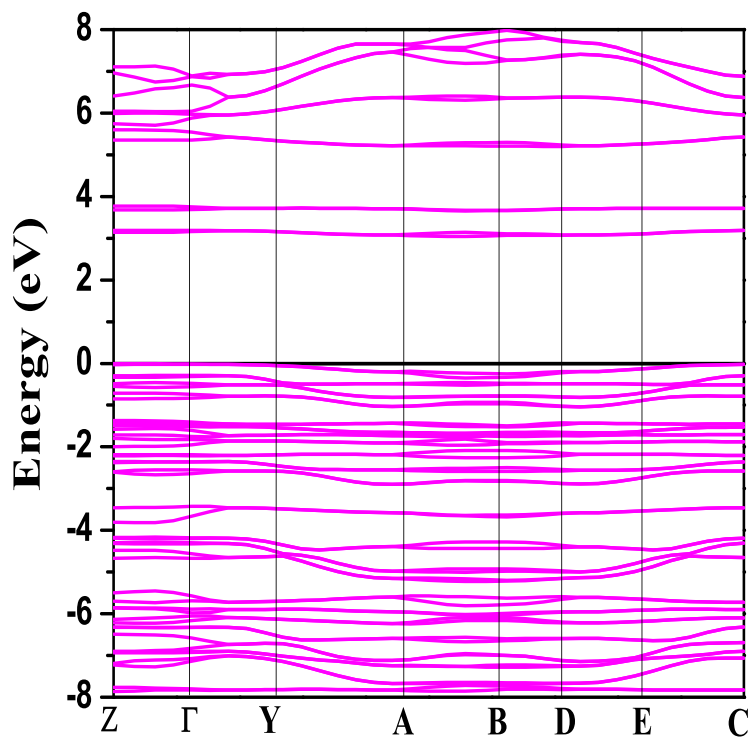}}
\subfigure[]{\includegraphics[width=120mm,height=80mm]{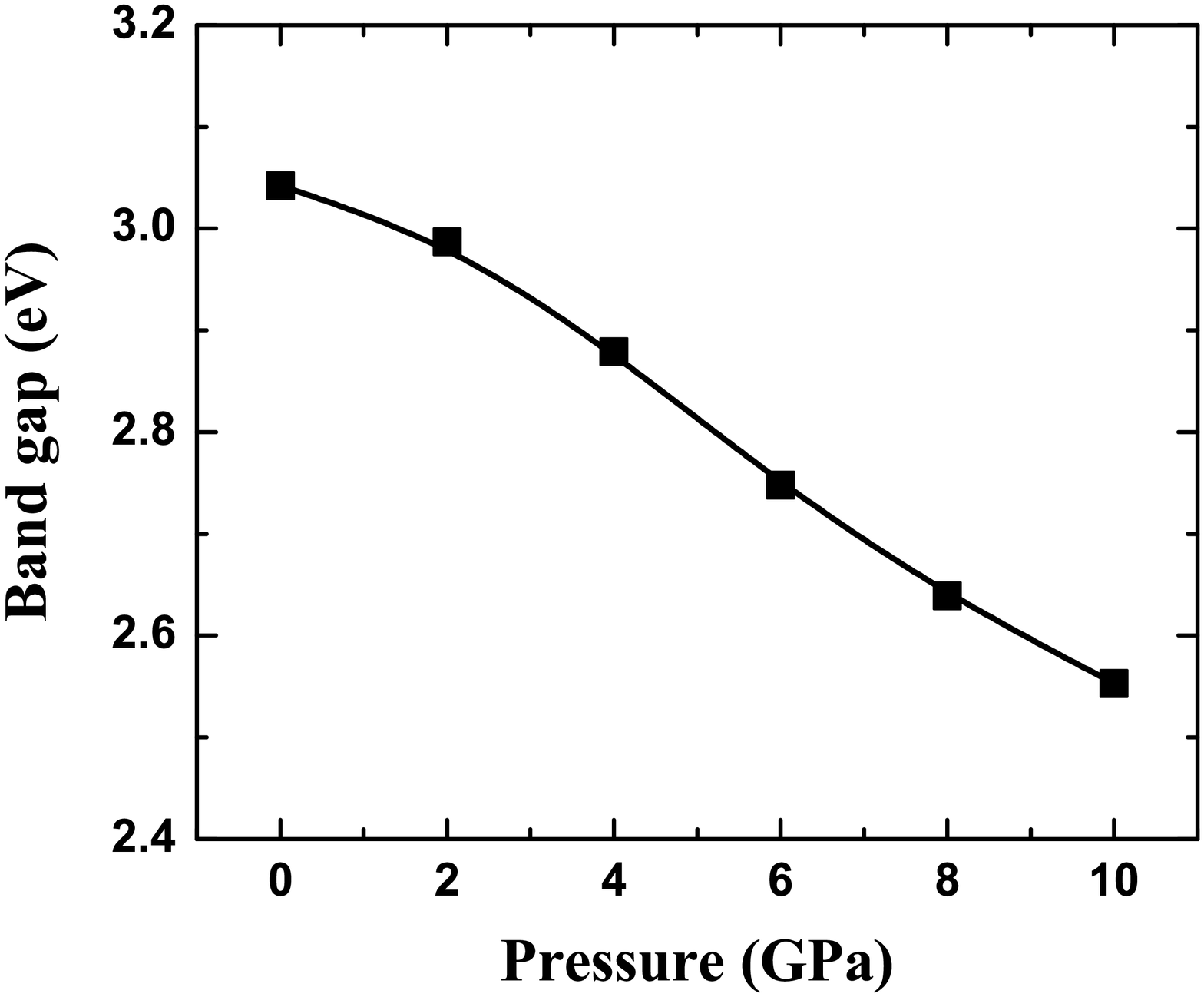}}\\

\caption{(Colour online) Electronic band structure of G-MNAT calculated within PBE+G06 functional (a). Variation of electronic band gap with pressure (b).}
\end{center}
\end{figure}
%
\clearpage
\newpage
\begin{table}[ht]
\caption{ Ground state structural properties of crystalline guanidinium 2-methyl-5-nitraminotetrazolate (G-MNAT) calculated using various exchange-correlation functionals. Experimental data have been taken from Fendt et. al \cite{Fendt}}
\begin{tabular}{ccccccccc} \hline
XC functional & a (\AA) & b (\AA) & c (\AA) & $\beta$$^0$ & $\rho$ & V (\AA$^3$)  \\ \hline \hline
CA-PZ & 3.5167&7.8685 &13.7515 &94.44 &1.778& 379.38\\
PBE & 4.7258 & 7.6277 & 15.1658 & 105.96 &1.284 & 525.61 \\
PBE+TS & 3.7123 & 8.0569 & 14.1369 & 95.43 &1.603 & 420.91 \\
PBE+G06 & 3.6111 & 8.0296 & 14.2334 & 95.35 & 1.642&  410.91\\
Expt & 3.6562 & 8.1552 &13.9458 & 95.91 &1.631  & 413.61 \\ \hline \hline
 & &B$_0$   & &  & B$_0$$^\prime$ & \\ \hline
PBE+G06 & & 15.3&   &  & 4.66 & \\ \hline \hline
\end{tabular}\\
\end{table}
\clearpage

\newpage
\begin{table}
\caption{The vibrational frequencies (cm$^{-1}$) of G-MNAT crystal calculated within PBE+G06 functional. Here LM represents lattice mode vibrations, $\omega$, $\delta$, and $\rho$ stands for wagging, bending,, and rocking of the modes respectively.}
\label{tab.1}
\begin{center}
\begin{tabular}{cccccc}\hline \hline
Freq&IrrRep&Mode Assignment&Freq&IrrRep&Mode Assignment \\ \hline
74.2& A & LM&188.2 &A& \\
78.3& A &&192.8 &A&$\omega$ NH$_2$ \\
83.9& B && 195.3&B& \\
87.7& A&& 212.2&B&$\omega$ NH$_2$, $\rho$ CH$_3$ \\
94.8& B& & 218.5&B& \\
99.7& A& & 223.5&A& \\
106.3& B& & 240.4&A&\\
108.5& A& & 244.1&B&  \\
116.1& B& & 256.6&A&\\
118.4 & A& & 257.5& B&\\
133.5& B& & 348.1&B&\\
137.5& A& & 354.6&A& \\
141.9& B& & 383.7&A&$\rho$ NH$_2$, $\delta$ N=N-N \\
145.9& A& &  384.5&B&\\
146.9& B& & 425.5&B&$\delta$ NH$_2$  \\
152.9& A& & 437.7&A& \\
154.9& B& & 445.8&A& \\
167.3& B& & 447.2&B& \\
168.5 & A& & 465.8&B&\\
174.2 & B& & 468.1&A&$\omega$ NH$_2$, $\omega$ CH$_3$, $\delta$ N=N-N \\
178.5 & A& & &\\\hline
\end{tabular}
\end{center}
\end{table}
\clearpage
\newpage
\begin{table}
\caption{The vibrational frequencies (cm$^{-1}$) of G-MNAT crystal calculated within PBE+G06 functional. The symbols $\nu$, $\omega$, $\delta$, and $\rho$ stands for stretch, wagging, bending, and rocking of the modes respectively.}
\label{tab.1}
\begin{center}
\begin{tabular}{cccccc}\hline \hline
Freq&IrrRep&Mode Assignment&Freq&IrrRep&Mode Assignment \\ \hline
523.8& A&$\omega$ NH$_2$ &1177.1&B &$\nu$ N-N, $\omega$ NH$_2$\\%
532.8&B& &1180.4&A & \\
557.6&A& & 1189.6&B&  \\
567.4&B&& 1190.7&A&$\delta$ ring\\
585.8& A && 1198.8&B &\\
600.1& B && 1199.1&A&  \\
606.6& A &&  1235.9&B&$\delta$ N-CH$_3$, $\delta$ C=N-N \\
633.8&B && 1236.1&A & \\
643.3& B &$\delta$ N-C-N, $\delta$ C-N-N & 1297.1&B&$\delta$ NO$_2$, $\delta$ N-CH$_3$\\
650.9& B&$\delta$ N=N-N, $\delta$ N-N-C & 1303.1& A&$\nu$ NO$_2$, $\nu$ N=N \\
657.3& A& & 1315.4&B&\\
683.8& A & $\delta$ N-C-N, $\omega$ NH$_2$& 1318.1&A& \\
687.2& B && 1330.5&A&$\omega$ NH$_2$,  $\delta$ C-N-N \\
704.9&B& &  1332.1&B& \\
715.2& A&$\omega$ N=C & 1350.6 &A &$\delta$ N-C-H, $\nu$ C-N \\
715.8& A && 1357.9&B&\\
716.7 & B&$\nu$ N-C-N,$\delta$ NO$_2$ & 1408.2&B&\\
727.5& B && 1421.9&A&$\delta$ C-N-N, $\omega$ NH$_2$  \\
727.6 & A& & 1435.7&B&\\
729.1& A& & 1445.4&A& \\
730.6&B& & 1451.3&A&$\delta$ N-C-H, $\nu$ C-N \\
774.2& A&$\omega$ NH$_2$ &1452.8&B&\\%
797.6&B& &1463.8&A &\\
848.1&B&$\nu$ N-C-N,$\nu$ C=N,$\nu$ N-CH$_3$, & 1464.3&B&  \\
854.4&A&& 1554.9&B&$\nu$ C=N \\
864.4& A && 1555.8&A& \\
864.6& B && 1574.9&A &$\delta$ NH$_2$, $\delta$ C=N-N \\
977.9& A &$\nu$ C-N, $\rho$ NH$_2$&  1577.5&B&\\
978.8&B && 1608.1&A& \\
995.9& A && 1656.1&B&$\delta$ NH$_2$, $\nu$ C-N, $\delta$ C-N-H \\
996.9& B& & 1666.6& B&\\
1010.3& A& $\nu$ N-N, $\nu$ NO$_2$, $\omega$ C-H& 1672.6&A&\\
1010.5& B & $\delta$ N-C-N, $\omega$ NH$_2$& 1678.9&B& \\
1027.8& A && 1687.9&A& $\nu$ C=N \\
1030.6&B& &  &&\\
1066.5& B& & & & \\
1069.4& A &$\nu$ N-N, $\omega$ C-NH$_2$ & &&\\
1078.6& B& & && \\
1091.5& A && & &\\
1109.7& A& & &&\\
1110.7&B& & && \\\hline
\end{tabular}
\end{center}
\end{table}
\clearpage
\newpage
\begin{table}
\caption{The stretching vibrational frequencies (cm$^{-1}$) of C-H and N-H modes of G-MNAT crystal calculated within PBE+G06 functional.}
\label{tab.1}
\begin{center}
\begin{tabular}{ccc}\hline \hline
Freq&IrrRep&Mode Assignment \\ \hline
2963.5&A&$\nu$$_{sym}$ C-H\\
 2965.1&B & \\
 3051.2&B&\\
 3051.9&A& \\
  3056.9&B &\\
   3062.8&A&\\
 3077.3&B& \\
 3078.1&A&\\
 3140.9& B&$\nu$$_{asy}$ C-H \\
 3143.8&A& \\
 3180.8 & B& \\
 3196.7 & A & \\
 3242.8 & B &$\nu$$_{sym}$ N-H \\
 3246.9 & A & \\
 3279.9 & A & \\
 3316.7 & B & $\nu$$_{asy}$ N-H \\
 3473.4 & A & \\
 3483.3 & B & \\ \hline \hline
\end{tabular}
\end{center}
\end{table}


\begin{thebibliography}{}
\bibitem{Agrawal}
J. P. Agrawal, High Energy Materials, Propellants, Explosives and Pyrotechniques, Wiley-VCH Verlag, Germany.
\bibitem{Klapotke}
Thomas M Klap\"{o}tke, Chemistry of High Energy Materials, (2011) Walter de Gruyter GmbH and Co. KG, Berlin/New York.
\bibitem{Akhavan}
J. Akhavan, The Chemistry of Explosives,(2011) Royal Society of Chemistry: Cambridge, UK.
\bibitem{Fair}
 H. D. Fair, and R. F. Walker, Energetic Materials, vol 1, (1977) Plenum Press, New York.
\bibitem{Talawar}
D. M. Badgujar, M. B. Talawar, S. N. Asthana, P. P. Mahulikar, J. Hazard. Mater 151 (2008) 289-305.
\bibitem{Sikder}
M. B. Talawar, R. Sivabalan, T. Mukundan, H. Muthurajan, A. K. Sikder, B. R. Gandhe and A. S. Rao, J. Hazard. Mater 161 (2009) 589-607.
\bibitem{Eremets}
M. I. Eremets, A. G. Gavriliuk, I. A.Trojan, D. A. Dzivenko, R. Boehler, Nat. Mater 3 (2004) 558.
\bibitem{Fendt}
T. Fendt, N. Fischer, Thomas M Klap\"{o}tke, J. Stierstorfer, Inorg. Chem 50 (2011) 1447.
\bibitem{Oleyn}
I. I. Oleynik, M. Conroy, S. V. Zybin, L. Zhang, A. C. van Duin, W. A. Goddard III, C. T. White, Shock Compression of Condensed Matter CP 845 (2005) 573.
\bibitem{YMGupta}
W. F. Perger, S. Vutukuri, Z. A. Dreger, Y. M. Gupta, K. Flurchick, Chem. Phys. Lett 422 (2006) 397-401.
\bibitem{Perger}
W. F. Perger, J. Zhao, J. M. Winey, Y. M. Gupta, Chem. Phys. Lett 428 (2006) 394-399.
\bibitem{Byrd}
E. F. C. Byrd, B. M. Rice, J. Phys. Chem. C 111 (2007) 2787.
\bibitem{Chaba}
E. F. C. Byrd, G. E. Scuseria, C. F. Chabalowski, J. Phys. Chem. B 108 (2004) 13100.
\bibitem{Rice}
D. C. Sorescu, B. M. Rice, J. Phys. Chem. C 114 (2010) 6734.
\bibitem{Grimme}
S. Grimme, J. Comp. Chem. 27 (2006) 1787.
\bibitem{TS}
A. Tkatchenko, M. Scheffler, Phys. Rev. Lett 102 (2009) 073005.
\bibitem{Xiao}
W. Zhu, H. Xiao, Struct. Chem 21 (2010) 657-665.
%
%
%
%
\bibitem{Segall}
 M. Segall, P. Lindan, M. Probert, C. Pickard, P. Hasnip, S. Clark, M. J. Payne, J. Phys.: Cond. Matt. 14 (2002) 2717.

 \bibitem{Vanderbilt}
 D. Vanderbilt, Phys. Rev. B 41 (1990) 7892.

\bibitem{Ceperley}
 D. M. Ceperley, B. J. Alder, Phys. Rev. Lett 45(1980) 566.

\bibitem{PPerdew}
J. P. Perdew, A. Zunger, Phys. Rev. B 23 (1981) 5048.

\bibitem{Perdew}
J. P. Perdew, K. Burke, M. Ernzerhof, Phys. Rev. lett 77 (1996) 3865.

\bibitem{Monkhorst}
H. J. Monkhorst, J.  Pack, Phys. Rev. B 13 (1976) 5188.

\bibitem{Conroy}
M. W. Conroy, I. I. Oleynik, S. V. Zybin, C. T. White, Phys. Rev. B 77 (2008) 094107.

\bibitem{Stevens}
J. J. Haycraft, L. L. Stevens, C. J. Eckhardt, J. Chem. Phys 124 (2006) 024712.

\bibitem{Chen}
H. L. Cui, G. F. Ji, X. Chen, W. Zhu, F. Zhao, Y. Wen, D. Wei, J. Phys. Chem. A 114 (2010) 1082-1092.

\bibitem{McWilliams}
R. S. McWilliams, Y. Kadry, M. F. Mahmood, A. F. Goncharov, J. C. Jenkins, J. Chem. Phys 137 (2012) 054501.

%

\bibitem{Gonze}
X. Gonze, Phys. Rev. B 55 (1997) 10337.

\bibitem{Refson}
K. Refson, P. R. Tulip, S. J. Clarke, Phys. Rev. B 73 (2006) 155114.

\bibitem{Bar}
S. Baroni, S. de Gironcoli, A. Dal Corso and P. Giannozzi, Rev. Mod. Phys 73 (2001) 515.

\end{thebibliography}
\end{document}